\documentclass[aps,twocolumn]{revtex4}
\usepackage{epsfig}
\usepackage{psfrag}
\newcommand{\be}{\begin{equation}}
\newcommand{\ee}{\end{equation}}
\newcommand{\bea}{\begin{eqnarray}}
\newcommand{\eea}{\end{eqnarray}}
\newcommand{\ba}{\begin{array}}
\newcommand{\ea}{\end{array}}
\newcommand{\E}{{\sf E}}

\begin{document}

\title{On Generalized Sch\"urmann Entropy Estimators}

\author{Peter Grassberger}
\affiliation{JSC, J\"ulich Research Center, D-52425 J\"ulich, Germany}

\date{\today}

\begin{abstract}
We present a new class of estimators of Shannon entropy for severely undersampled 
discrete distributions. It is based on a generalization of an estimator 
proposed by T. Schuermann, which itself is a generalization of an estimator
proposed by myself in arXiv:physics/0307138. For a special set of parameters 
they are completely free of bias and have a finite variance, something with is 
widely believed to be impossible. We present also
detailed numerical tests where we compare them with other recent estimators
and with exact results, and point out a clash with Bayesian estimators for 
mutual information.
\end{abstract}

\maketitle

It is well known that estimating (Shannon) entropies from finite samples
is not trivial. If one naively replaces the probability $p_i$ to be in 
``box" $i$ by the observed frequency, $p_i \approx n_i/N$, statistical
fluctuations tend to make the distribution look less uniform, which 
leads to an underestimation of the entropy. There have been numerous
proposals on how to estimate and eliminate this bias 
\cite{miller,harris,herzel,grass88,schmitt,wolpert,poschel,panzeri,schuermann,strong,holste,nemenman,paninski,grassberger2003,schuermann2004,vu,bonachela,hausser,wolpert2013,chao,archer,hernandez}. 
Some make quite strong assumptions \cite{schmitt,poschel}, others use Bayesian
methods \cite{wolpert,holste,nemenman,wolpert2013,archer,hernandez}. As pointed 
out in \cite{grass88,grassberger2003,paninski,bonachela}, one can devise 
estimators with arbitrarily small bias (for sufficiently large $N$ and fixed 
$p_i$), but these will then have very large statistical errors. As conjectured 
in \cite{grass88,grassberger2003,paninski,schuermann2004,bonachela} the variance
of any estimator whose bias vanishes will have a diverging variance. 

Another wide spread believe is that Bayesian entropy estimators cannot be 
outperformed by non-Bayesian ones for severely undersampled cases. The problem
with Bayesian estimators is of course that they depend on a good choice of 
prior distributions, which is not always easy, and they tend to be slow. One 
positive feature of a non-Bayesian estimator proposed in \cite{grassberger2003}
is that it is extremely fast, since it works precisely like the `naive' (or 
maximum-likelihood) estimator, except that the logarithms used there are 
replaced by a function $G_n$ defined on integers, which can be precomputed by
means of a simple recursion. While the estimator of \cite{grassberger2003} 
seems in general to be a reasonable compromise between bias and variance, it 
was shown in \cite{schuermann2004} that it can be improved -- as far as bias 
is concerned, at the cost of increased variance -- by generalizing $G_n$ to
a one-parameter family of functions $G_n(a)$.

In the present letter we will show that the Grassberger-Schuermann approach
\cite{grassberger2003,schuermann2004} can be further improved by using different
functions $G_n(a_i)$ for each different realization $i$ of the random variable.
Indeed, the $a_i$ can be chosen such that the estimator is completely free of
bias and yet has a finite variance -- although, to be honest, the optimal
parameters $a_i$ can be found only, if the exact distribution is known (in which
case also the entropy can be computed exactly). We will show that -- even if 
the exact optimal $a_i$ are not known -- the new estimator can reduce the 
bias very much, without inducing unduly large variances, provided the 
distribution is sufficiently much undersampled (sic!).

We shall test the proposed estimator numerically with simple examples, where we 
shall produce bias-free entropy estimates e.g. from pairs of ternary variables, 
something which to my knowledge is not possible with any Bayesian method. We
shall also use it for estimating mutual information (MI) in cases where one of the 
two variables is binary, and the other one can take very many values. In the 
limit of severe undersampling and of no obvious regularity in the true 
probabilities, MI cannot be estimated unambiguously. In that limit, the present
algorithm seems to choose systematically a different outcome from Bayesian 
methods, for reasons that are not yet clear.

In the following we shall use the notation of \cite{grassberger2003}. As in
this reference, we consider $M > 1$ ``boxes" (states, possible experimental
outcomes, ...) and $N > 1$ points (samples, events, particles) distributed
randomly and independently into the boxes. We assume that each box has weight
$p_i$ ($i=1,\ldots M$) with $\sum_i p_i = 1$. Thus each box $i$ will contain
a random number $n_i$ of points, with $\E[n_i] = p_iN$. Their joint
distribution is multinomial,
\be
   P(n_1,n_2,\ldots n_M;N) = N! \prod_{i=1}^M\frac{p_i^{n_i}}{n_i!},    \label{miltinomi}
\ee
while the marginal distribution in box $i$ is binomial,
\be
   P(n_i;p_i,N) = {N\choose n_i} p_i^{n_i} (1-p_i)^{N-n_i} .    \label{binomi}
\ee

Our aim is to estimate the entropy, 
\be
   H = -\sum_{i=1}^M p_i \ln p_i = \ln N - {1\over N}\sum_{i=1}^M z_i \ln z_i,
\ee
with $z_i \equiv \E[n_i] = p_iN$,
from an observation of the numbers $\{n_i\}$ (in the following, all 
entropies are measured in ``natural units", not in bits). The estimator 
$\hat{H}(n_1,\ldots n_M)$ will of course have both statistical 
errors and a bias, i.e. if we repeat this experiment, the average
of $\hat{H}$ will in general not be equal to $H$,
\be
   \Delta [\hat{H}] \equiv \E[\hat{H}] - H \neq 0,
\ee
as will also be its variance ${\rm Var}[\hat{H}]$. Notice that for 
computing $\E[\hat{H}]$ we need only Eq.(2), not the full multinomial
distribution of Eq.(1). But if we want to compute this variance, we need 
in addition the joint marginal distribution in two boxes,
\bea
   P(n_i,n_j;p_i,p_j,N) & = & \frac{N!}{n_i!n_j!(N-n_i-n_j)!} \;\times \\
                 & &  p_i^{n_i} p_j^{n_j}(1-p_i-p_j)^{N-n_i-n_j}, \nonumber
                 \label{bibinomi}
\eea
in order to compute the covariances between different boxes.
Notice that these covariances were not taken into account in 
\cite{paninski,bonachela}, whence the variance estimations in these papers 
are at best approximate.

In the following, we shall mostly be interested in the limit 
$N\to\infty, M\to\infty$, with $M/N$ (the average number of points 
per box) finite and small. In this limit the variance will
go to zero (because essentially one averages over many boxes), but 
the bias will remain finite. The binomial distribution, Eq.(\ref{binomi}),
can be replaced then by a Poisson distribution
\be
   P_{\rm Poisson}(n_i;z_i) = {z_i^{n_i}\over n_i!} e^{-z_i}.         \label{poiss}
\ee
More generally, we shall also be interested
in the case of large but finite $N$, where also the variance is positive, 
and we will discuss the balance between demanding minimal bias versus 
minimal variance.

Indeed it is well known that the {\it naive} (or `maximum-likelihood') 
estimator, obtained by assuming $z_i = n_i$ without fluctuations,
\be
   \hat{H}_{\rm naive} = \ln N - {1\over N}\sum_{i=1}^M n_i \ln n_i,
\ee
is negatively biased, $\Delta \hat{H}_{\rm naive} < 0$.

In order to estimate $H$, we have to estimate $p_i\ln p_i$ or equivalently 
$z_i \ln z_i$ for each $i$. Since the distribution of $n_i$ depends, 
according to Eq.(\ref{binomi}), on $z_i$ only, we can make the rather 
general ansatz \cite{grass88,grassberger2003}
for the estimator
\be
    \widehat{z_i \ln z_i} = n_i\phi(n_i)
   \label{zi}
\ee
with a yet unknown function $\phi(n)$.
Notice that $\hat{H}$ becomes with this ansatz a sum over strictly positive 
values of $n_i$. Effectively this means that we have assumed that observing 
an outcome $n_i=0$ does not give any information: If $n_i=0$, we do not 
know whether this is because of statistical fluctuations or because $p_i=0$ 
for that particular $i$.

The resulting entropy estimator is then \cite{grassberger2003}
\be
   \hat{H}_\phi = \ln N - {M\over N} \overline{n\phi(n)}  \label{8}
\ee
with the overbar indicating an average over all boxes,
\be
   \overline{n\phi(n)} = {1\over M} \sum_{i=1}^M n_i \phi(n_i).
\ee
Its bias is
\be
   \Delta H_\phi = {M\over N} (\overline{z\ln z} -\overline{\E_{N,z}[{n\phi(n)}]}).
   \label{dHphi}
\ee
with 
\be
    \E_{N,z}[f_n] = \sum_{n = 1}^\infty f_n P_{\rm binom}(n;p=z/N,N). 
\ee
being the expectation value for a typical box (in the following we shall
suppress the box index $i$ to simplify notation, wherever this makes sense).

In the following, $\psi(x) = d\ln \Gamma(x)/dx$ is the digamma function, and
\be
   E_1(x)=\Gamma(0,x) = \int_1^\infty {e^{-xt}\over t} dt
\ee
is an exponential integral (Ref.\cite{abramow}, paragraph 5.1.4). It was shown 
in \cite{grassberger2003} that
\be
  \E_{N,z}[n\psi(n)] = z\ln z + z[\psi(N)-\ln N]  
                   +  z\int_0^{1-z/N}{x^{N-1}dx\over 1-x} \;,
		    \label{psi}
\ee
which simplifies in the Poisson limit ($N\to\infty$, $z$ fixed) to 
\be
  \E_{N,z}[n\psi(n)] \to z\ln z + z E_1(z)\;. \label{psi_poisson}
\ee

Eqs.(\ref{psi}) and (\ref{psi_poisson}) will be the starting points of all further 
analysis. In \cite{grassberger2003} it was proposed to re-write Eq.(\ref{psi_poisson})
as 
\be
  \E_{N,z}[nG_n] \to z\ln z + z E_1(2z)\;, \label{G}
\ee
where 
\be
   G_n = \psi(n) + (-1)^n\int_0^1{x^{n-1}\over x+1} dx.      
\ee
The advantages are that $G_n$ can be evaluated very easily by recursion (here 
$\gamma=0.57721...$ is the Euler-Mascheroni constant),
$ G_1 =G_2 = -\gamma - \ln 2,\;\; G_{2n+1} = G_{2n},\;$ and $ G_{2n+2} = G_{2n}+{2\over 2n+1}$,
and neglecting the second term, $z E_1(2z)$, gives an excellent approximation 
unless $z$ is exceedingly small, i.e. unless the numbers of points per box 
are very small so that the distribution is very severely undersampled. Thus 
the entropy estimator proposed in \cite{grassberger2003} was simply
\be
   \hat{H}_G = \ln N - {1\over N} \sum_{i=1}^M n_i G_{n_i}.
    \label{twoz}
\ee
Furthermore, since $z E_1(2z)$ is positive definite, neglecting it gives a negative bias
in $\hat{H}_G$, and one can show rigorously that this bias is smaller than that
of \cite{miller,herzel}.

The easiest way to understand the Schuermann class of estimators \cite{schuermann2004} 
is to define, instead of $G_n$, a one-parameter family of functions
\be
   G_n(a) = \psi(n) + (-1)^n\int_0^a{x^{n-1}\over x+1} dx.     \label{Gan}
\ee
Notice that $G_n(1)=G_n$ and $G_n(0) = \psi(n)$.

Let us first discuss the somewhat easier Poissonian limit, where
\bea
  \E_{N,z}&&\!\!\!\!\!\!\![n(G_n(a) - \psi(n))] =\nonumber \\
               && = \sum_{n=1}^\infty (-1)^n P_{\rm Poisson}(n,z) \int_0^a {x^{n-1}\over x+1} dx \nonumber \\
	       && = -ze^{-z} \int_0^a {dx\over x+1} e^{-xz} \nonumber \\
	       && = -z(E_1(z)-E_1((1+a)z)),    
	             \label{Ga_poisson}
\eea
which gives 
\be
    \E_{N,z}[nG_n(a)] = z \ln z + zE_1((1+a)z).
\ee
Using -- to achieve optimality -- different parameters $a_i$ for different boxes, and 
neglecting the right hand side of Eq.(\ref{Ga_poisson}), we obtain finally
\be
   \hat{H}_{\rm Schuermann} = \ln N - {1\over N} \sum_{i=1}^M n_i G_{n_i}(a_i) \qquad {\rm (Poisson)}.
    \label{schuer}
\ee
The reason why the r.h.s. of Eq.(\ref{Ga_poisson}) can be neglected for sufficiently 
large $a_i$ is simply that $0 < E_1(bz) < \exp(-bz)$ for any real $b>0$.

For the general binomial case the algebra is a bit more involved.
By somewhat tedious but straightforward algebra one finds that
\bea 
  \E_{N,z}&&\!\!\!\!\!\!\![n(G_n(a) - \psi(n))] =\nonumber \\ 
               && =\sum_{n=1}^\infty (-1)^n {N\choose n} p^n(1-p)^{N-n} \int_0^a {x^{n-1}\over x+1} dx \nonumber \\
	       && =-pN\int_0^a{dx\over x+1} \sum_{n=1}^\infty {N-1\choose n-1}(-px)^{n-1} (1-p)^{N-n} \nonumber \\
	       && =-pN \int_0^a{dx\over x+1} (1-p-px)^{N-1}\nonumber \\
	       && =-z \int_0^a{dx\over x+1}[ 1-\frac{(1+x)z}{N}]^{N-1}.
\eea
One immediately checks that this reduces, in the limit ($N\to\infty$, $z$ fixed), to 
Eq.(\ref{Ga_poisson}). On the other hand, by substituting 
\be
    x \to t = 1-\frac{(1+x)z}{N}
\ee
in the integral, we obtain 
\be
  \E_{N,z}[n(G_n(a) - \psi(n))] = -z \int_{1-(1+a)z/N}^{1-z/N}\frac{t^{N-1}dt}{1-t}.
\ee
Finally, by combining with Eq.(\ref{psi}), we find \cite{schuermann2004}
\bea
  \E_{N,z}[n(G_n(a)] &=& z \ln z + z[\psi(N)-\ln N] + \nonumber \\
	  &+& z\int_0^{1-(1+a)z/N}{x^{N-1}dx\over 1-x} 
	    \label{Ga-binom}
 \eea
and 
\be
   \hat{H}_{\rm opt} = \psi(N) - {1\over N} \sum_{i=1}^M n_i G_{n_i}(a_i), \qquad {\rm (binomial)}
       \label{H-sch}
\ee
with a correction term whose bias vanishes when the integration range on the r.h.s.
of Eq.(\ref{Ga-binom}) is zero. Notice that we use here, in general, a different 
parameter $a_i$ for each box $i$. In \cite{schuermann2004} one single parameter $a$
was used, which is why we call our method a generalized Schuermann estimator.

This is a remarkable result, as it shows that the analytic correction to the naive 
estimator, as given by Eqs.(19) and (27), become exact when for each box $i$
\be
   a_i \to a_i^* \equiv \frac{1-p_i}{p_i}.    \label{ai}
\ee
When all box weights are small, $p_i \ll 1$ for all $i$, then these bias-optimal
values of $a_i$ are very large. But for two boxes with $p_1=p_2=1/2$, e.g., 
the bias vanishes already for $a_1=a_2=1$, i.e. for the estimator of 
Grassberger \cite{grassberger2003}! 

In order to test the latter, we drew $10^8$ triplets of random bits (i.e., 
$N=3$, $p_0=p_1=1/2$), and estimated $\hat{H}_{\rm naive}$ and 
$\hat{H}_G$ for each triplet. From these we computed averages and 
variances, with the results $\hat{H}_{\rm naive} = 0.68867(4)$ bits and 
$\hat{H}_G = 0.99995(4)$ bits. We should stress that the latter requires 
the precise form of Eq.(\ref{H-sch}) to be used, with $\psi(N)$ neither 
replaced by $\ln N$ nor by $G_N$.

Since there is no free lunch, there must of course be some problems in
the limit when parameters $a_i$ are chosen nearly bias-optimal. One 
problem is that one cannot, in general, choose $a_i$ according to 
Eq.(\ref{ai}), because the $p_i$ are unknown. In addition, it is 
in this limit (and more generally when $a_i>>1$) that variances 
blow up. In order to see this, we have to discuss in more detail
the properties of the functions $G_n(a)$.

According to Eq.(\ref{Gan}), $G_n(a)$ is a sum of two terms, both of which
can be computed, for all positive integer $n$, by recursion. The digamma
function $\psi(n)$ satisfies
\be
    \psi(1) = -\gamma, \quad \psi(n+1) = \psi(n) + 1/n .
\ee
Let us denote the integral in Eq.(\ref{Gan}) as $g_n(a)$. It satisfies 
the recursion
\be
    g_1(a) = -\ln(1+a), \quad g_{n+1}(a) = g_n(a) +(-a)^n/n.
\ee
Thus, while $\psi(n)$ is monotonic and slowly increasing, $g_n(a)$ has 
alternating sign and increases, for $a>1$, exponentially with $n$. As a 
consequence, also $G_n(a)$ is non-monotonic and diverges exponentially
with $n$, whenever $a>1$. Therefore an estimator like $\hat{H}_{\rm opt}$
gets, unless all $n_i$ are very small, increasingly large contributions 
of alternating signs. As a result variances will blow up, unless one is 
very careful to keep a balance between bias and variance.

\begin{figure}
  \begin{center}
  \psfig{file=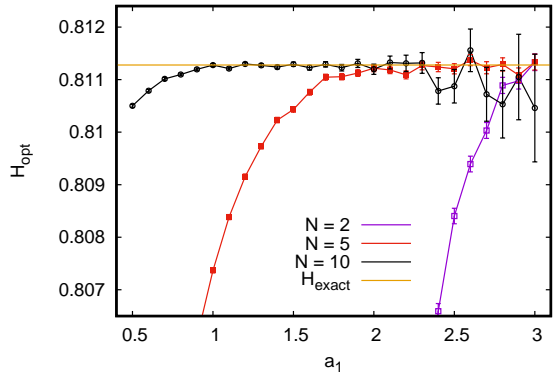,width=8.5cm, angle=0}
  \vglue -25pt
  \caption{Estimated entropies (in bits) of $N$-tuples of i.i.d. random 
     binary variables with $p_0= 3/4$ 
     and $p_1 = 1/4$, using the optimized estimator $\hat{H}_{\rm opt}$
     defined in Eq.(\ref{H-sch}). The parameter $a_0$ was kept fixed at its 
     optimal value $a_0 = 1/3$, while $a_1$ is varied in view of possible 
     problems with the variances, and is plotted on the horizontal axis.
     For each $N$ and each value of $a_1$, $10^8$ tuples were drawn. The 
     exact entropy for $p_0= 3/4$ and $p_1 = 1/4$ is $0.811278\ldots$ bits,
     and is indicated by the horizontal straight line.}
    \label{binary2}
  \end{center}
  \vglue -10pt
\end{figure}

\begin{figure}
  \begin{center}
  \psfig{file=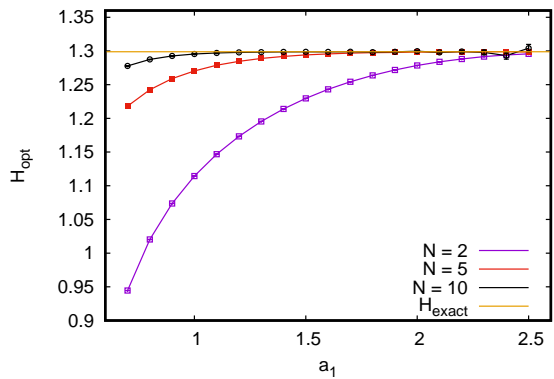,width=8.5cm, angle=0}
  \vglue -25pt
  \caption{Estimated entropies (in bits) of $N$-tuples of i.i.d. random
     ternary variables with $p_0= 0.625, p_1 = 0.25$, and $p_2 = 0.125$, 
     using the optimized estimator $\hat{H}_{\rm opt}$ defined in 
     Eq.(\ref{H-sch}). The parameter $a_0$ was kept fixed at its optimal 
     value $a_0^* = 0.6$, while $a_1$ and $a_2$ varied in view of possible
     problems with the variances. More precisely, we used $a_2 = 1+4 (a_1-1)$, 
     so that the data end at the bias-free values $a_1^*=2.5$ and $a_2^*=7.0$
     For each $N$ and each value of $a_1$, $10^8$ tuples were drawn. The
     exact entropy is $1.29879\ldots$ bits,
     and is indicated by the horizontal straight line.}
    \label{ternary}
  \end{center}
  \vglue -10pt
\end{figure}

To illustrate this we drew tuples of i.i.d. binary variables 
$\{s_1,\ldots s_N\}$ with $p_0 = 3/4$ and $p_1 = 1/4$. For $a_0$ we chose
$a_0 = a_0^* = 1/3$, because this should minimize the bias and should 
not create problems with the variance. We should expect such problems, 
however, if we would take $a_1 = a_1^* = 3$, although this would reduce the 
bias to zero. Indeed we found for $N=100$ that the variance of the estimator 
exploded for all practical purposes as soon as $a_1 > 1.4$, while the 
results were optimal for $0.5 < a_1 \leq 1$ (bias and statistical error
were both $< 10^{-5}$ for $10^8$ tuples). On the other hand, for pairs
($N=2$) we had to use much larger values of $a_1$ for optimality, and $a_1=3$ 
gave indeed the best results (see Fig.1). A similar plot for ternary variables
is shown in Fig.2, where we see again that the bias-optimal values gave 
estimates with zero bias and acceptable variance for the most undersampled 
case $N=2$.

The message to be learned from this is that we should always keep all
$a_i$ sufficiently small that $a_i^{n_i} \leq O(1)$ for the observed values 
of $n_i$.

Finally, we apply our estimator to two problems of mutual information
(MI) estimation discussed in \cite{hernandez} (actually, the problems were 
originally proposed by previous authors, but we shall compare our results
mainly to those in \cite{hernandez}. In each of these problems there are two
discrete random variables: $X$ has many (several thousand) possible values, 
while $Y$ is binary. Moreover, the marginal distribution of $Y$ is uniform,
$p(y=0) = p(y=1)=1/2)$, while the $X$-distributions are highly non-uniform.
Finally -- and that is crucial -- the joint distributions show no obvious
regularities. 

\begin{figure}
  \begin{center}
  \psfig{file=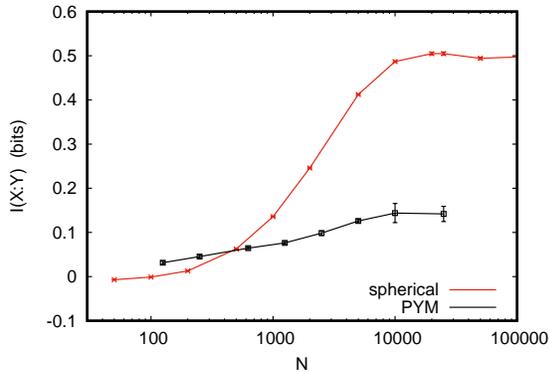,width=8.0cm, angle=0}
  \vglue -25pt
  \caption{Estimated mutual informations (in bits) of $N$-tuples of i.i.d. random
     subsamples from two distributions given in \cite{hernandez}. The 
	  data for "PYM", originally due to \cite{schwartz}, consist of 
	  250,000 pairs $(x,y)$ with binary $y$ with $p(y=0)=p(y=1)=1/2$,
	  and $x$ being uniformly distributed over 4096 values. Thus each
	  $x-$value is realized $\approx 60$ times, and we classify them
	  into 5 classes depending on the associated $y-$values: (i) very 
	  heavily biased towards $y=1$, (ii) moderately biased towards $y=1$,
	  (iii) $y-$neutral, (iv) moderately biased towards $y=0$, and (v) 
	  heavily biased towards $y=0$. When we estimated conditional entropies
	  $H(Y|X)$ for randomly drawn subsamples, we kept this classification
	  and choose $a_y$ accordingly: For class (iii) we used $a_0=a_1=1$, 
	  for class (ii) we used $a_1=1, a_0=4$, for class (i) we used
	  $a_1=1, a_0=7$, for class (iv) we used  $a_1=4, a_0=1$, and finally 
	  for class (v) we used $a_1=7, a_0=1$. The data for "spherical",
	  originally due to \cite{archer}, consist of 50,000 $(x,y)$ pairs.
	  Here, $Y$ is again binary with $p(y=0)=p(y=1)=1/2$, but $X$ is 
	  highly non-uniformly distributed over $\approx 4000$ values. Again 
	  we classified these values as 
	  $y-$neutral or heavily / moderately biased towards or against $y=0$
	  and used this classification to choose values of $a_y$ accordingly.}
    \label{damian}
  \end{center}
  \vglue -10pt
\end{figure}

The MI is estimated as $I(X:Y) = H(Y) -H(Y|X)$. Since $H(Y) = 1$ bit, the 
problem essentially burns down to estimate the conditional probabilities 
$p(y|x)$. The data are given in terms of a large number of i.i.d. sampled
pairs $(x,y)$ (250,000 pairs for problem I, called `PYM' in the following,
and 50,000 pairs for problem II, called `spherical' in the following).
The task is to draw random subsamples of size $N$, to estimate the MI
from each subsample, and to calculate averages and statistical widths from
these estimates.

Results are shown in Fig.~3. For large $N$ our data agree perfectly with 
those in \cite{hernandez} and in the previous papers cited in \cite{hernandez}. 
But while the MI estimates in these previous papers all increase with 
decreasing $N$, and those in \cite{hernandez} stay essential constant
(as we would expect, since a good entropy estimator should not depend on
$N$, and conditional entropies should decrease with $N$ for not so good
estimators), our estimated MI decreases to zero for small $N$.

This looks at first sight like a failure of our method, but it is not.
As we said, the joint distributions show no regularities. For small $N$
most values of $X$ will show up at most once, and if we write the sequence
of $y-$values in a typical tuple, it will look like a perfectly random
binary string. The modeler knows that it actually is not random, because
there are correlations between $X$ and $Y$. But no algorithm can know 
this, and any good algorithm should conclude that $H(Y|X) = H(Y) = 1$ bit.
Why, then was this not found in the previous analyses? In all these,
Bayesian estimators were used. If the priors used in these estimators 
were chosen in view of the special structures in the data (which are, 
as we should stress again, not visible from the data, as long as these are 
severely undersampled!), then the 
algorithms can make of course use of these structures and avoid the 
conclusion that $H(Y|X) = 1$ bit.

In conclusion, we have given an entropy estimator with zero bias and finite 
variance. It builds on an estimator by Schuermann \cite{schuermann2004} 
which itself is a generalization of \cite{grassberger2003}. It involves
a real-valued parameter for each possible realization of the 
random variable, and bias is reduced to zero by choosing
these parameters properly. But this would require that we know already
the distribution, which is of course not the case. Nevertheless we can 
reduce the bias very much for severely undersampled cases. In cases with
moderate undersampling, choosing these zero-bias parameters would give very
large variances and would thus be useless. Nevertheless, by judicious 
parameter choices we can obtain extremely good entropy estimates. Finding
good parameters is non-trivial, but is made less difficult by the fact that 
the method is very fast.

Finally, we pointed out that Bayesian methods which have been very popular
in this field have the danger of choosing ``too good'' priors, i.e. choosing 
priors which are not justified by the data themselves and are thus misleading,
although both the bias and the observed variances seem to be small.

I thank Thomas Schuermann for numerous discussions, and Dami\'an Hern\'andez 
for both discussions and for sending me the data for Fig.~3.

\end{document}